\documentclass[12pt]{iopart}
\usepackage{iopams}  

\usepackage{color}
\usepackage{amsthm}
\usepackage[dvips]{graphicx}

\theoremstyle{plain}
\newtheorem{theorem}{Theorem}[section]

\theoremstyle{definition}

\newtheorem{remark}{Remark}[section]

\newcommand{\bt}{\mbox{\boldmath$t$}}

\newcommand{\bq}{\mbox{\boldmath$q$}}
\newcommand{\bii}{\mbox{\boldmath$i$}}
\newcommand{\bj}{\mbox{\boldmath$j$}}
\newcommand{\bn}{\mbox{\boldmath$n$}}

\newcommand{\bear}[1]{\hspace{-.3em}\begin{array}{#1}}
\newcommand{\ear}{\end{array}}
\newcommand{\eqref}[1]{(\ref{#1})}
\newcommand{\blank}[1]{}

\let\cd\cdot
\let\vphi\varphi
\let\p\partial
\let\ds\displaystyle

\begin{document}

\title[On non-steady planar motions of fibre-reinforced fluids]{On non-steady planar motions of fibre-reinforced fluids. Geometry and integrable structure}

\author{D.K. Demskoi${}^1$, W.K. Schief${}^2$}

\address{${}^1$ School of Computing and Mathematics,\\ Charles Sturt University, NSW 2678, Australia \\[2mm]
${}^2$ School of Mathematics and Statistics,\\ University of New South Wales, Sydney NSW 2052, Australia}

\begin{abstract}
It is shown that the kinematic system describing planar non-steady  motions of ideal fibre-reinforced fluids may be reduced to a single two-dimensional third-order partial differential equation in which time enters parametrically. A procedure is established which maps steady motions to non-steady motions.
The resulting motions inherit their hidden integrable structure from the steady case. The formalism presented here also readily recovers the connection with the scattering problem of the modified Korteweg-de Vries hierarchy established in previous work.
\end{abstract}

\vspace{2pc}
\noindent{\it Keywords}: fibre-reinforced fluid, geometry, integrability
%\maketitle

\section{Introduction}
The mathematical theory of the deformation of fibre-reinforced materials has been set down in a monograph by Spencer \cite{spencer}. Resin matrix fibre-reinforced materials have abundant engineering applications, notably in the construction of robust, light, laminated shell structures with complicated geometries. The formation process of fibre-resin systems in which the resin matrix behaves as a viscous fluid is of particular industrial interest. In this connection, Hull et al.\ \cite{hull} adopted an ideal fibre-reinforced fluid model to describe fibre-resin systems in the important formation state. This model consists of an incompressible fluid which is inextensible along ``fibre'' lines which occupy the volume of the fluid by which they are convected. The presence of these privileged fibre orientations in the fluid imposes strong kinematic constraints on its admissible motions. Thus, if a generic fibre direction is characterised by a unit vector $\bt$ and~$\bq$ denotes the fluid velocity then the kinematic condition which encodes both the convection requirement and inextensibility of the fibres is given by
\begin{equation}
\left(\p_t+{\bq}\cd \nabla \right){\bt}=({\bt} \cd \nabla){\bq}.
\label{msys}
\end{equation}
This evolution equation is augmented by the continuity equation 
\begin{equation}
	\mathrm{div}\,{\bq}=0
	\label{divqzero1}
\end{equation}
which follows from the incompressibility of the fluid. An important implication of the kinematic system (\ref{msys}), (\ref{divqzero1}) is obtained by calculating the divergence of (\ref{msys}) and taking into account (\ref{divqzero1}). The resulting relation turns out to be
\begin{equation}
	\left(\p_t+{\bq}\cd \nabla\right) \mbox{div}\, {\bt}=0
	\label{theta_0}
\end{equation} 
which expresses the fact that
\begin{equation}
    \theta=\mbox{div}\,{\bt}
\end{equation} 
%i
is preserved along the particle lines. It is emphasised that, according to Spencer \cite{spencer2},
two-dimensional flows of an ideal fibre-reinforced fluid are privileged in that these are essentially determined by kinematic considerations. Therefore, pressure $p$ and tension $T$ in the fibre direction can always be determined such that the equations of motion are satisfied. 

As part of a programme which aims to identify hidden integrable structure in nonlinear continuum mechanics (see \cite{schiefjmp} and references therein), the kinematic equations \eqref{msys}, \eqref{divqzero1} have been investigated in detail. In \cite{schiefquart,schieftmf}, it has been demonstrated that, in the planar steady case, this pair of equations is reducible to a single nonlinear third-order partial differential equation which, in turn, admits an integrable reduction associated with the B\"acklund transformation for the celebrated sine-Gordon equation \cite{RogersSchief2002}. Moreover, by virtue of \eqref{theta_0}, it is evident that the case $\mathrm{div}\,\bt=\mathrm{const}$ may be of particular interest. Indeed, in \cite{murugesh}, non-steady planar motions subject to this condition have been shown to be encoded in the AKNS scattering problem for the modified Korteweg-de Vries (mKdV) hierarchy \cite{AblowitzClarkson1991}. Iterated Darboux transformations \cite{RogersSchief2002} have been used to construct explicitly flows of fibre-reinforced fluids within regions which are bounded by two parallel curves of non-trivial geometry with the fibres constituting generalised tractrices.

Guided by the previous work cited above, we here develop a formalism (Section~2) which is well suited for the treatment of the governing equations of non-steady planar motions of fibre-reinforced fluids. In Section 3, we adopt a canonical adapted coordinate system to prove that, remarkably, such motions are, without imposition of any constraints, captured by the same two-dimensional third-order partial differential equation as in the steady case, wherein time enters parametrically. The only difference is that the arbitrary function involved in this equation now depends not only on arc length along the fibres (as in the steady case) but also on the coordinate parametrising the lines orthogonal to the fibres. In Section 4, we then demonstrate how the parametric dependence on time may be exploited to map steady motions to non-steady motions to which we refer as {\it quasi-steady} motions. In particular, this implies that the integrable structure encoded in the steady case is inherited by the quasi-steady case. Furthermore, we explain how the mKdV connection observed when the fibre divergence is constant may be readily retrieved within the current formalism.

\section{Algebraic and geometric properties of the governing equations}

\subsection{Geometric decomposition}

It is convenient to parametrise the unit vector $\bt$ according to
\begin{equation}\label{A1}
\bt={\bii}\cos \varphi+{\bj}\sin \varphi,
\end{equation}
where ${\bii}$ and ${\bj}$ are the vectors of the standard Cartesian orthonormal basis.
Hence, the unit normal to the fibres adopts the form
\begin{equation}\label{A2}
\bn=-{\bii}\sin \varphi+{\bj}\cos \varphi.
\end{equation}
We denote the derivatives along the fibres and their orthogonal trajectories respectively by
\begin{equation}
		D_s = {\bt} \cd \nabla,\quad D_n= {\bn}\cd \nabla,
	\label{dsdnintro}
\end{equation}
which implies the decomposition
\begin{equation}
\nabla={\bt} D_s+{\bn} D_n.
\end{equation}
It is noted that the fibre divergence may be expressed in terms of the derivative in the normal direction as
\begin{equation}\label{thetaexpression}
\theta=D_n\vphi.
\end{equation}
The commutator relations for the operators (\ref{dsdnintro}) and $\p_t$ may be calculated directly and are given by
\begin{equation}
	\left[\p_t,D_s\,\right]=\vphi_tD_n,\quad
	\left[\p_t,D_n\,\right]=-\vphi_tD_s,\quad
	[D_n,D_s]=\kappa D_s+\theta D_n, 
	\label{com1}
\end{equation}
where 
\begin{equation}
\kappa=D_s\varphi = -\mathrm{div}\,\bn
\label{kappaintro}
\end{equation}
constitutes the curvature of the fibres. Using these relations, we can rewrite (\ref{msys}) in the form
\begin{equation}
\Big(\vphi_t +({\bq}\cd \nabla \vphi)\Big) {\bn}=D_s{\bq}.
	\label{msys2}
\end{equation}
Furthermore, we will exploit the fact that (\ref{msys}) may be formulated as
\begin{equation}
	\left[\p_t+{\bq} \cd \nabla,D_s\right]=0,
	\label{com3}
\end{equation}
which expresses the fact that the inextensible fibres are convected with the fluid.

It is convenient to resolve the velocity ${\bq}$ into components along the fibres and normal to them according to
\begin{equation}
{\bq}=v{\bt}+w{\bn}.
\end{equation}
The latter gives rise to the decomposition
\begin{equation}
{\bq}\cd \nabla=v D_s+w D_n.
\label{decomp}
\end{equation}
On substitution of (\ref{decomp}) into (\ref{com3}) and use of (\ref{com1}), we obtain the relations
\begin{equation}
\ds D_s w=\vphi_t+\theta w, \quad \ds D_s v=\kappa w.
\label{wsvs}
\end{equation}
Then, condition (\ref{divqzero1}) adopts the form
\begin{equation}
	D_s v+D_n w+v \theta-w \kappa=0.
	\label{divqzero}
\end{equation}
Thus, combination of (\ref{wsvs}), (\ref{divqzero}) produces the key system
\begin{equation}
D_s w=\vphi_t+ \theta w, \quad
D_n w=- \theta v, \quad
D_s v= \kappa w,
\label{sys3}
\end{equation}
which is equivalent to the original governing equations (\ref{msys}), (\ref{divqzero1}).

\subsection{Compatibility conditions}

We now investigate the compatibility of the system (\ref{sys3}). Firstly, on use of the commutator relation (cf.\ $(\ref{com1})_3$)
\begin{equation}
	[D_n,D_s]w=\kappa D_s w+\theta D_n w,
	\label{com_ns_w}
\end{equation}
the compatibility condition for the pair $(\ref{sys3})_1$, $(\ref{sys3})_2$ is seen to be
\begin{equation}
 D_n(\vphi_t + \theta w) + D_s(\theta v) = \kappa(\vphi_t + \theta w) - \theta^2v,
\end{equation}
which, by virtue of the commutator relation (cf.\ $(\ref{com1})_2$)
\begin{equation}
  [\p_t,D_n]\vphi = -\vphi_tD_s\vphi
\end{equation}
and the relations (\ref{thetaexpression}), (\ref{kappaintro}), simplify to
\begin{equation}
(\p_t+v D_s+w D_n)\theta=0,
\label{comp}
\end{equation}
thereby reproducing the important consequence (\ref{theta_0}) of the original governing equations.

In view of \eqref{comp}, it is natural to introduce an arc length parameter $\rho$ along the fibres which is convected with the fluid according to
\begin{equation}\label{rhointro}
  D_s\rho = 1,\quad (\p_t+v D_s+w D_n)\rho=0.
\end{equation}
The latter pair is compatible due to the commutativity \eqref{com3} of the convective derivative and $D_s$. We may therefore solve for the velocity component $v$ to obtain
\begin{equation}
	v=- wD_n\rho-\rho_t.
	\label{vexpr}
\end{equation}
Now, application of the commutator relations  (\ref{com1})$_{1,3}$ to $\rho$ results in  
\begin{equation}
 D_s\rho_t = - \vphi_tD_n\rho,\quad D_sD_n\rho=-\kappa - \theta D_n\rho
\end{equation}
so that substitution of (\ref{vexpr}) into the remaining relation (\ref{sys3})$_3$ reveals that the latter is satsified modulo (\ref{sys3})$_1$.

\subsection{Determination of the divergence $\theta$ and the velocity $\bq$}

It turns out convenient to introduce another (non-constant) quantity which is convected by the fluid but constant along the fibres, that is,
\begin{equation}\label{alphaintro}
  D_s\alpha = 0,\quad (\p_t+v D_s+w D_n)\alpha=0.
\end{equation}
It is important to stress that $\rho$ and $\alpha$ are functionally independent since $D_s\alpha=0$  but $D_s\rho=1$. Moreover, since the general solution of \eqref{comp} is an arbitrary function of two functionally independent particular solutions, we conclude that
\begin{equation}
  \theta = \theta(\rho,\alpha).
\end{equation}
Since $D_n\alpha=0$ would imply that $\alpha=\mathrm{const}$, we may solve \eqref{alphaintro}$_2$ for $w$ to obtain
\begin{equation}\label{wexpr}
  w = -\frac{\p_t\alpha}{D_n\alpha}.
\end{equation}
Application of the commutator relations (\ref{com1}) to $\alpha$ yields
\begin{equation}\label{dmitry}
 D_s\alpha_t = -\vphi_tD_n\alpha,\quad D_n\alpha_t = \p_tD_n\alpha,\quad D_sD_n\alpha=-\theta D_n\alpha
\end{equation}
so that substitution of $v$ and $w$ as given by \eqref{vexpr}, \eqref{wexpr} into \eqref{sys3}$_{1,2}$ and subsequent evaluation shows that \eqref{sys3}$_1$ is identically satisfied and \eqref{sys3}$_2$ may be formulated as a vanishing Jacobian condition, namely
\begin{equation}
  J_{snt}(\rho,\alpha,D_n\alpha) = 0.
\end{equation}
Here, the Jacobian is calculated with respect to the operators $D_s$, $D_n$ and $\p_t$. Since $\rho$ and $\alpha$ are functionally independent, the above condition states that
\begin{equation}\label{star}
D_n\alpha = M(\rho,\alpha).
\end{equation}
Hence, relation \eqref{dmitry}$_3$ becomes
\begin{equation}
 M_{\rho}=-\theta M
\label{DsM}
\end{equation}
so that
\begin{equation}
 \theta = -\frac{M_\rho}{M}.
 \label{thetaformula}
\end{equation}

In summary, for an arbitrary function $M(\rho,\alpha), $ where $\rho$ and $\alpha$ are functions of the spatial coordinates and time subject to the conditions  $D_s\rho=1$, $D_s\alpha=0$ and \eqref{star}, the quantity $\theta$ defined by (\ref{thetaformula}) and the velocity components \eqref{vexpr}, \eqref{wexpr} obey the kinematic equations \eqref{sys3}.

\section{Derivation of the governing partial differential equation}\label{intrcoordsec}

\subsection{An adapted coordinate system}

We now introduce a spatially orthogonal coordinate system of the type
\begin{equation}
	x = x(s,n,\tau),\quad y = y(s, n,\tau),\quad t=\tau.
	\label{newvars}
\end{equation}
We choose the coordinates $s$ and $n$ in such a manner that they parametrise the fibres and their orthogonal trajectories respectively.
Hence, the metric of the plane adopts the form
\begin{equation}
	g=S^2 ds^2+N^2 dn^2,
	\label{planemetric}
\end{equation}
that is, for any fixed time $t$, the coordinate lines are orthogonal.
Then, the partial derivatives associated with the new coordinates are related to the directional derivatives $D_s$ and $D_n$ via
\begin{equation}
	\frac{1}{S}\frac{\p}{\p s}=D_s,\quad \frac{1}{N}\frac{\p}{\p n}=D_n. 
	\label{partialdirect}
\end{equation}

\begin{remark}
\label{remark2} It is important to note that the functions $S$ and $N$ are defined up to (local) scaling factors of the form $\hat{S}(s,\tau)$ and $\hat{N}(n,\tau)$ respectively, corresponding to 
reparametrisations of the $s$- and $n$-lines.
\end{remark}

Substitution of (\ref{partialdirect}) into the commutator relation $(\ref{com1})_3$ allows us to relate the quantities $S, N,$ and $\vphi$. Indeed, $(\ref{com1})_3$ translates into
\begin{equation}\label{NS}
N_s=\theta N S,\quad  S_n=-\kappa N S
\end{equation}
so that, by virtue of the relations
\begin{equation}
  \kappa = D_s\vphi = \frac{\vphi_s}{S},\quad \theta = D_n\vphi = \frac{\vphi_n}{N},
\end{equation}
we may rewrite the pair \eqref{NS} as
\begin{equation}\label{phinphis}
N_s=\vphi_n S  , \quad S_n= -\vphi_s N.
\end{equation}
Relation (\ref{rhointro})$_1$ now shows that
\begin{equation}
  S = \rho_s
\end{equation}
and, hence, we obtain
\begin{equation}
	\frac{N_s}{N}=\theta S=-\frac{M_\rho}{M}\rho_s
	\label{NsNthetaS}
\end{equation}
with $\theta$ given by \eqref{thetaformula} so that
\begin{equation}
N=\frac{\hat{c}(n,\tau)}{M}.
\end{equation}
Due to Remark \ref{remark2}, we may set $\hat{c}=1$ without loss of generality and, hence,
\begin{equation}\label{Nformula}
 M=\frac{1}{N},
\end{equation}
which reveals that $N=N(\rho,\alpha)$. Accordingly, relations (\ref{phinphis}) may be formulated as 
\begin{equation}
	\vphi_n=N_\rho,\quad \vphi_s=-\frac{\rho_{ns}}{N}. 
	\label{flaphi}
\end{equation}
The compatibility condition $\vphi_{ns}=\vphi_{sn}$ then results in the {\it two-dimensional} partial differential equation
\begin{equation}
\left(\frac{\rho_{sn}}{N}\right)_n+N_{\rho\rho}\rho_s=0.
\label{eqlast}
\end{equation}
Finally, relation $(\ref{partialdirect})_2$ applied to $\alpha$ together with \eqref{star} yields $\alpha_n=1$ and, hence, 
\begin{equation}
\alpha=n+f(\tau)
\end{equation}
since $D_s\alpha = 0$. Then, by appropriately re-parametrising the $n$-lines, it may be achieved that the function $f(\tau)$ vanishes and $N$ assumes the form
\begin{equation}
	N=N(\rho,\alpha)=N(\rho,n).
	\label{Nrhon}
\end{equation}

\begin{remark}
Remarkably, in \cite{schiefquart}, it has been shown that steady planar motions of fibre-reinforced fluids are governed by a differential equation identical to (\ref{eqlast}) except that, in that case, $N$ is a function of $\rho$ only as discussed in Section 4.
\end{remark}

\subsection{Equations for the position and velocity vectors}\label{randv}
It follows from (\ref{partialdirect}) and \eqref{A1}, \eqref{A2} that
\begin{equation}
	x_s=\rho_s\cos \vphi,\quad\ x_n=-N\sin \vphi, \quad y_s=\rho_s\sin \vphi, \quad y_n=N\cos\vphi
	\label{xyparam}
\end{equation}
which are, by construction, compatible modulo the pair \eqref{flaphi}. In order to recover the temporal dependence, we need to find the coefficients of the expansion
\begin{equation}
	\p_t=\p_\tau+\beta\p_s+\gamma\p_n.
	\label{dtop}
\end{equation}
The relations $x_t=y_t=0$ lead to
\begin{equation}
x_\tau+\beta x_s+\gamma x_n=0, \quad
y_\tau+\beta y_s+\gamma y_n=0.
\end{equation}
On substituting (\ref{xyparam}), we obtain the expressions
\begin{equation}
	\beta=-\frac{x_\tau\cos\vphi+y_\tau\sin\vphi }{\rho_s},\quad \gamma=\frac{x_\tau\sin\vphi-y_\tau\cos\vphi}{N}.
\label{betagamma}
\end{equation}
Hence, once $\rho$ is known, the coordinates $x$ and $y$ are obtained by integrating (\ref{xyparam}). Relations (\ref{vexpr}) and (\ref{wexpr}) then determine the components of the velocity vector. Accordingly, we can formulate the key result of this paper which summarises the analysis presented in this section.

\begin{theorem}\label{theorem1}
Given a solution $(\rho,N(\rho,n))$ of the partial differential equation
\begin{equation}\label{thirdgeneric}
\left(\frac{\rho_{sn}}{N}\right)_n+N_{\rho\rho}\rho_s=0,
\end{equation}
wherein $\rho$ may depend parametrically on $\tau$, a solution of the governing equations (\ref{msys}), (\ref{divqzero1}) is given by 
\begin{equation}
{\bt}=(\cos\vphi, \sin\vphi),\quad {\bq}=v\, (\cos\vphi, \sin\vphi)+w \,(-\sin\vphi, \cos\vphi),
\end{equation}
where the angle $\vphi$ is determined by the compatible system
\begin{equation}
\vphi_s=-\frac{\rho_{ns}}{N},\quad\vphi_n=N_{\rho}
\end{equation}
and the velocity components $v$ and $w$ read
\begin{equation}
	v = x_\tau\cos \vphi +y_\tau \sin \vphi-\rho_\tau, \quad w = -x_\tau \sin \vphi+y_\tau \cos \vphi
	\label{velocityvwgeneric}
\end{equation}
with the parametrisation of the Eulerian coordinates being obtained via integration of the compatible system
\begin{equation}
	x_s=\rho_s\cos \vphi,\quad\ x_n=-N\sin \vphi, \quad y_s=\rho_s\sin \vphi, \quad y_n=N\cos\vphi
\end{equation}
and $t=\tau$. At any instant $t$, the one-parameter family of fibres labelled by $n$ is parametrised by $s$ according to $(x,y)(s,n,\tau=t)$.
\end{theorem}

\begin{remark}
It is remarkable that Theorem \ref{theorem1} comprises a partial differential equation and quadratures which do not explicitly involve the temporal coordinate $\tau$. It is therefore instructive to verify this theorem directly by expressing $(\p_x,\p_y,\p_t)$ in terms of $(\p_s,\p_n,\p_{\tau})$ via \eqref{xyparam}, \eqref{dtop} and \eqref{betagamma} and evaluating the governing equations (\ref{msys}), (\ref{divqzero1}) modulo the relations stated in the theorem. Of course, for the adapted coordinate system to exist, one makes the assumption that $|\p(x,y,t)/\p(s,n,\tau)|=\rho_sN \neq 0$. The verification may be considerably simplified by exploiting the expression
\begin{equation}
  \p_t + \bq\cdot\nabla = \p_{\tau} - \frac{\rho_{\tau}}{\rho_s}\p_s
\end{equation}
for the convective derivative, which is a consequence of the relations \eqref{dtop}, \eqref{betagamma} and \eqref{velocityvwgeneric}. It is noted that the above identity immediately confirms that the quantities $\rho$ and $\alpha=n$ are convected with the fluid.
\end{remark}

\section{(Quasi-)steady motions. Integrability}

\subsection{Steady motions}

The class of steady planar motions is retrieved by first assuming that $N$ is a function of $\rho$ only and then applying a simple reparametrisation of the $n$-lines corresponding to 
\begin{equation}
  \p_\tau\rightarrow \p_\tau + c\p_n,\quad c=\rm{const}
\end{equation} 
with the derivatives $\p_s$ and $\p_n$ being unchanged. Since we have assumed that $N=N(\rho)$, Theorem \ref{theorem1} subject to $N_n=0$ does not change except that the velocity components now read
\begin{equation}
 v = x_\tau\cos \vphi +y_\tau \sin \vphi-\rho_\tau - c\rho_n, \quad w = -x_\tau \sin \vphi+y_\tau \cos \vphi + cN.
\end{equation}
If we now demand that all functions be independent of $\tau$, the velocity components simplify to
\begin{equation}
  v = -c\rho_n,\quad w = cN
\end{equation}
and Theorem \ref{theorem1} coincides with that established in \cite{schiefquart} in the context of steady planar motions.

\subsection{Quasi-steady motions}

Motions for which the fibre divergence $\theta$ is a function of arc length $\rho$ only are algebraically privileged since the associated constraint $\theta = {(\ln N)}_{\rho} = \Theta(\rho)$ leads to a separable function
\begin{equation}
  N = \nu(n)\mathcal{N}(\rho).
\end{equation}
Since the function $\nu(n)$ may be scaled to unity by a suitable re-parametrisation of the $n$-lines, we refer to motions of this type as {\it quasi-steady} motions for the following important reason. In light of the previous subsection, any (sufficiently non-trivial) steady planar motion may be boosted to obtain non-steady motions by assuming that the constants of integration in the associated solution $\rho$ of \eqref{thirdgeneric} with $N=N(\rho)$ constitute arbitrary functions of $\tau$ and applying Theorem~\ref{theorem1}. This procedure may be applied to, for instance, the steady planar motions generated by the soliton solutions of the integrable reduction~\cite{schieftmf}
\begin{equation}\label{sine}
 \rho_s = \sinh\sigma,\quad \sigma_n = \sin\rho,\quad N = \sin\rho
\end{equation}
of the third-order equation \eqref{thirdgeneric}, leading to quasi-steady motions governed by ``modulated'' soliton solutions of \eqref{thirdgeneric} with the modulation driven by, for example, time-dependent spectral parameters. It is noted that this integrability connection is due to the fact that the pair \eqref{sine}$_{1,2}$ constitutes (on complexification) the classical B\"acklund equations for the integrable sine-Gordon equation \cite{RogersSchief2002}. In fact, it is easy to see that the quantity $\rho + \rmi\sigma$ satisfies the sine-Gordon equation
\begin{equation}
  {(\rho + \rmi\sigma)}_{sn} = \sin (\rho + \rmi\sigma).
\end{equation}
The exact (mathematical and physical) nature of this ``boosting algorithm'' together with exact solutions are considered elsewhere.

\subsection{The mKdV hierarchy}

Another application of soliton theory \cite{AblowitzClarkson1991} is based on the discovery made in \cite{murugesh} that non-steady motions subject to constant fibre divergence admit integrable structure associated with the scattering problem \cite{AblowitzClarkson1991}
\begin{equation}\label{scattering}
 \left(\bear{c}y_1\\ y_2\ear\right)_s = \frac{1}{2}\left(\bear{cc} \lambda\, & -u\\ u\, & -\lambda\ear\right)\left(\bear{c}y_1\\ y_2\ear\right)
\end{equation}
of the modified Korteweg-de Vries (mKdV) equation
\begin{equation}\label{mkdv}
 u_{\tau} = u_{sss} + \frac{3}{2}u^2u_s
\end{equation}
and its associated hierarchy. In the terminology adopted here, these motions evidently constitute special quasi-steady motions. In fact, since $\theta=1$, without loss of generality, and $\theta = {(\ln N)}_{\rho}$, in general, we deduce that
\begin{equation}\label{Nmkdv}
 N = \lambda^{-1}e^\rho,
\end{equation}
where, for convenience, the multiplicative factor $\nu(n)$ has been scaled to a constant as described above.

The integrability of the case \eqref{Nmkdv} is readily established within the present approach. In fact, this integrable connection provides an explicit example of the relevance of the parametric dependence on $\tau$ of our formalim. To this end, we first rewrite the third-order equation \eqref{thirdgeneric} as the coupled system
\begin{equation}\label{coupled}
  \rho_{sn} = \phi N,\quad \phi_n = -N_\rho\rho_s.
\end{equation}
Then, it is readily verified that, in the case \eqref{Nmkdv}, this system admits the first integral
\begin{equation}\label{integral}
  \phi^2 + \rho_s^2 = I(s,\tau).
\end{equation}
The latter may be parametrised according to
\begin{equation}\label{rho}
  \rho_s = \lambda \sin\sigma,\quad\phi = \lambda\cos\sigma
\end{equation}
so that the coupled system \eqref{coupled} reduces to
\begin{equation}\label{sigma}
  \sigma_n = \lambda^{-1}e^{\rho}.
\end{equation}
Here, we have used a suitable re-parametrisation of the $s$-lines to obtain $I=\lambda^2$.
Hence, the associated planar motions encapsulated in Theorem \ref{theorem1} are governed by the pair \eqref{rho}$_1$, \eqref{sigma}. 

We now eliminate $\rho$ between \eqref{rho}$_1$ and \eqref{sigma} and integrate to obtain
\begin{equation}\label{sineode}
  \sigma_s + \lambda\cos\sigma = u(s,\tau),
\end{equation}
where $u$ is a function of integration. In practice, explicit integration of this first-order ordinary differential equation may not be achieved for a generic function $u$. However, it turns out that infnite sequences of ``potentials'' $u$ for which integration is possible may be generated iteratively by means of Darboux transformations \cite{RogersSchief2002}. In order to verify this assertion, we make the substitution
\begin{equation}
  \sigma = 2\arctan\chi - \frac{\pi}{2}
\end{equation}
which transforms \eqref{sineode} into the Riccati equation
\begin{equation}
  \chi_s = \frac{u}{2}(1 + \chi^2) - \lambda\chi.
\end{equation}
Linearisation via
\begin{equation}
 \chi = \frac{y_2}{y_1}
\end{equation}
then indeed leads to the scattering problem \eqref{scattering} associated with the mKdV hierarchy, where the parameter $\lambda$ plays the role of the spectral parameter.

\begin{figure}
 \centerline{\includegraphics[width=0.7\textwidth]{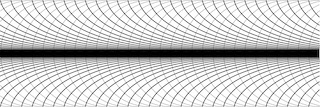}}
 \caption{Fibre distribution and orthogonal trajectories associated with a
straight base curve.}
\label{straight}
\end{figure}

As established in \cite{murugesh}, the fibre distributions of non-steady motions subject to $\theta=1$ are confined to a strip bounded by two curves which are parallel and at unit distance from a given base curve. The fibres constitute the generalised tractrices associated with the base curve and their orthogonal trajectories are circles of unit radius, which is consistent with the fact that $\theta$ constitutes the curvature of the $n$-lines. The seed potential $u=0$ corresponds to a (moving) straight base curve with associated fibre distribution displayed in Figure \ref{straight}.\newpage

\begin{figure}
 \centerline{\includegraphics[width=0.7\textwidth]{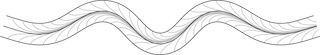}}
 \vspace{5mm}
 \centerline{\includegraphics[width=0.7\textwidth]{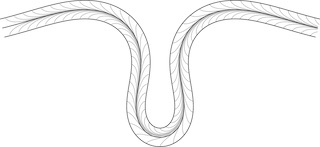}}
 \caption{Two fibre distributions associated with 
 breather potentials. The black curves represent the boundaries which contain the fluid flow and the base curves generating the fibre tractrices.}
\label{breather}
\end{figure}

Application of a Darboux transformation to the scattering problem \eqref{scattering} produces a one-soliton potential which generates non-physical motions since the fibres intersect. However, a second application of the Darboux transformation leads to a breather potential if the two B\"acklund parameters are complex conjugates. Associated fibre distributions are shown in Figure \ref{breather} for two different choices of the B\"acklund parameters. 
Details of this purely algebraic solution generation procedure based on the Sym-Tafel formula \cite{Sym1985} may be found in \cite{murugesh}. It should be mentioned that a link between the mKdV hierarchy and the motion of curves on the plane has been noted previously (see, e.g., \cite{GoldsteinPetrich1991}). In the current context, this is closely related to the motion of the base curve. However, the motion of the fibre distributions considered here is different.

\subsection{The Liouville equation}

We conclude by looking at the case \eqref{Nmkdv} from a different angle. Thus, if we introduce the complex quantity $\Omega = \rho + \rmi\sigma$ then the pair \eqref{rho}$_1$, \eqref{sigma} adopts the form
\begin{equation}\label{backlund}\fl
  \left(\frac{\Omega + \bar{\Omega}}{2}\right)_s = -\rmi\lambda\sinh\left(\frac{\Omega-\bar{\Omega}}{2}\right),\quad \left(\frac{\Omega - \bar{\Omega}}{2}\right)_n = \rmi\lambda^{-1}\exp\left(\frac{\Omega+\bar{\Omega}}{2}\right).
\end{equation}
Accordingly, $\bar{\Omega}$ constitutes a B\"acklund transform of $\Omega$ with $\rmi\lambda$ representing the B\"acklund parameter \cite{RogersSchief2002}. Indeed, if we temporarily set aside the fact that $\bar{\Omega}$ is the complex conjugate of $\Omega$ and regard the pair \eqref{backlund} as a system for $\bar{\Omega}$ then the compatibility condition $\bar{\Omega}_{sn} = \bar{\Omega}_{ns}$ is satisfied if and only if $\Omega$ is a solution of the Liouville equation
\begin{equation}
  \Omega_{sn} = e^{\Omega}.
\end{equation}
Hence, if $\Omega$ is a solution of the Liouville equation then, for reasons of symmetry,  $\bar{\Omega}$ determined by the compatible pair \eqref{backlund} constitutes another solution of the Liouville equation. In addition, one now imposes the admissible constraint that the B\"acklund transform $\bar{\Omega}$ is indeed the complex conjugate of the seed $\Omega$. Thus, it has been demonstrated that the $(s,n)$-dependence of non-steady motions subject to constant divergence $\theta$ is governed by the Liouville-type system \eqref{rho}$_1$, \eqref{sigma}, while the $(s,\tau)$-dependence is encoded in, for instance, the mKdV equation \eqref{mkdv}.

\section{Conclusion}

In previous work, the governing equations of planar motions of fibre-reinforced fluids have been analysed both algebraically and geometrically in important special cases. In particular, integrable structure residing in these equations has been identified. In this paper, we have demonstrated that one may develop a formalism which deals with the general case without imposition of any constraints. Moreover, we have shown how the previously discussed cases naturally arise and how their integrable structure may be retrieved and extended in the current context. The latter may now be exploited to generate and analyse large classes of non-steady motions. The results of this investigation will be presented elsewhere.

\end{document}